\documentclass[12pt,preprint,ams, nofootinbib]{revtex4}
\usepackage{amsmath}
\usepackage{amsfonts}
\usepackage{amsthm}
\usepackage{amssymb}
\usepackage{graphicx}
\usepackage{epsfig}
\usepackage{hyperref}
\usepackage{amscd}
\usepackage{xcolor}
\usepackage{enumitem}
\usepackage{thmtools, thm-restate}
\usepackage{graphics,epsfig,placeins,subfig,wrapfig}
\usepackage{appendix}

\def\be{\begin{equation}}
\def\ee{\end{equation}}

\def\bea{\begin{eqnarray}}
\def\eea{\end{eqnarray}}

\def\baa{\begin{align}}
\def\eaa{\end{align}}

\theoremstyle{definition}

\theoremstyle{Identity}

\begin{document}
\preprint{APS/123-QED}
\title{$C^3$ matching conditions for anisotropic fluids}

\author{Antonio  C. Guti\'errez-Pi\~{n}eres}
\email{acgutier@uis.edu.co}
\affiliation{Escuela de F\'\i sica, Universidad Industrial de Santander, CP 680002,  Bucaramanga, Colombia}
\affiliation{Instituto de Ciencias Nucleares, Universidad Nacional Aut\'onoma de M\'exico, AP 70543, Ciudad de M\'exico 04510,  Mexico }

\author{Hernando Quevedo}
\email{quevedo@nucleares.unam.mx}
\affiliation{Instituto de Ciencias Nucleares, Universidad Nacional Aut\'onoma de M\'exico, AP 70543, Ciudad de M\'exico 04510,  Mexico }
\affiliation{Dipartimento di Fisica and ICRA, Universit\`a di Roma ``Sapienza", I-00185, Roma, Italy}
\affiliation{Institute of Experimental and Theoretical Physics, 
	Al-Farabi Kazakh National University, Almaty 050040, Kazakhstan}

 \begin{abstract} 
The $C^3$  approach is an invariant formalism that  utilizes  the eigenvalues of the Riemann curvature tensor to match spacetimes across a specific matching surface. We apply this approach to match an anisotropic fluid with an exterior vacuum solution, including the case in which discontinuities appear on the matching surface.
As a particular example, a class of analytic solutions, which describe the gravitational field of realistic 
neutron stars, is matched to the exterior Schwarzschild spacetime.  
 \end{abstract}


\maketitle

\section{Introduction}
\label{sec:int}
To describe the gravitational field of compact objects in Einstein's theory, it is necessary to consider separately the interior and the exterior fields. 
For each of them, a solution of Einstein's equations should exist that satisfies the physical requirements of a compact object. 
The geometric background of Einstein's theory demands that the geometric properties of the spacetime be well-defined, 
implying that it is necessary to match the two solutions along a matching surface. 
The $C^3$ matching formalism was developed precisely to investigate this problem. 

In general relativity, the matching problem has been the subject of active research  for almost a century since Darmois published in 1927 his celebrated matching conditions \cite{darmois1927equations}. 
Darmois method  demands that the first and second fundamental forms be continuous along the surface of matching.
Other matching methods have been proposed \cite{lichnerowicz1955theories,bonnor1981junction,lichnerowicz1939problemes,oppenheimer1939continued,sharif2011dynamics,giang2009velocity,wiltshire2003slowly}, which, similar to Darmois' procedure, essentially impose conditions on the second derivatives of the spacetime metric ($C^2$ matching). 
A practical difficulty related to the $C^2$ matching is that it requires the use of ``admissible" coordinates that are not always available \cite{israel1966singular}. Israel proposed a generalization of the $C^2$ approach, using explicitly Darmois matching conditions, in which the matching surface is replaced by a thin shell when the continuity conditions are not satisfied \cite{israel1966singular}. Other matching procedures based on the use of the metric and curvature tensors have been shown to be equivalent to the Darmois conditions \cite{lake2017revisiting}.

The $C^3$ matching \cite{quevedo2012matching} is an alternative approach that does not depend on the choice of coordinates because it is based on the use of scalar quantities, namely, the eigenvalues of the Riemann curvature tensor. The main idea of this approach is simple. It is demanded that along the matching surface, the eigenvalues of the interior solution coincide with the eigenvalues of the exterior solution. Moreover, the local extrema of the  eigenvalues of the exterior spacetime are used to fix the location of the matching surface. Furthermore,  in the $C^3$ approach, it is possible to handle the case of surface discontinuities by using the thin-shell formalism proposed by Israel \cite{israel1966singular,gutierrez2022darmois}.
 
 In a previous work (\cite{gutierrez2019c3}), we studied the $C^3$ matching conditions for asymptotically 
 flat spacetimes in the framework of relativistic astrophysics. We concluded that in perfect-fluid spacetimes with spherical symmetry, the density and pressure 
 must vanish at the matching surface. Recently (\cite{gutierrez2022darmois}), we applied the $C^3$ matching approach to different perfect-fluid solutions of Einstein equations, which are considered as interior spacetimes that can be matched with the exterior Schwarzschild solution,  but contain  discontinuities on the matching surface. To handle this case, we proposed a  generalization of the  $C^3$ matching procedure. It consists essentially in demanding that the  matching surface $\Sigma$ is described by a solution of Einstein equations with a physically meaningful energy–moment tensor, which describes the matter inside 
a boundary shell located on the matching surface $\Sigma$. 

In this work, we show that it is possible to extend the previously $C^3$ matching approach to scenarios in which solutions of Einstein equations, which could be considered
appropriate interior candidates to be matched with the exterior Schwarzschild solution, describe fluids with anisotropic pressures. We present here  a $C^3$ matching approach, which specializes in the above anisotropic case, focusing  on static and spherically symmetric spacetimes. As a practical example of this method, a set of analytic solutions, describing the gravitational field of neutron stars, is matched with an exterior Schwarzschild spacetime. 

This work is organized as follows. Section \ref{sec:C3} provides an overview of the $C^3$ matching approach. In Sec. \ref{sec:sph}, we discuss the $C^3$ matching approach 
in the context of anisotropic fluids. Furthermore, in Sec. \ref{sec:C3NS}, we specifically focus on applying the $C^3$ matching procedure to neutron stars.
Finally, in Sec. \ref{sec:CR} general conclusions are presented.

\section{$C^3$ matching  approach} 
\label{sec:C3}
 
 The $C^3$ matching procedure is based on the analysis of the behavior of the Riemann 
curvature eigenvalues. Here, we employ the Cartan formalism of differential forms and local orthonormal tetrads to determine these 
eigenvalues. A local orthonormal tetrad is the simplest and most natural choice for an observer in order to perform local measurements of time, 
space, and gravity. So, let us choose the local orthonormal tetrad  $\vartheta^a$, $a=0,..., 3$ such that 
 \begin{align}
\mathbf{\cal G}= g_{\mu\nu}  \operatorname{d} x^\mu\otimes  \operatorname{d} x^\nu= \eta_{ab}\vartheta^a\otimes\vartheta^b\ ,
\end{align}
with $\eta_{ab}={\rm diag}(-1,1,1,1)$, and $\vartheta^a = e^a_{\ \mu}dx^\mu$. The first and second Cartan equations
\begin{align}
 \operatorname{d} \vartheta^a = - \omega^a_{\ b }\wedge \vartheta^b\ ,  & \\
\Omega^a_{\ b} =\operatorname{d}\omega^a_{\ b} + \omega^a_{ \ c} \wedge \omega^c_{\ b} & 
                          = \frac{1}{2}  R^a_{\ bcd} \vartheta^c\wedge\vartheta^d
\end{align}
allow us to compute the components of the Riemann curvature tensor $R_{abcd}$  in the local orthonormal frame $\vartheta^a$. 
Moreover, we define the Ricci tensor and the scalar curvature as  $R_{ab} = R^{c}_{\ acb} $ and $R= R^{a}_{\ a} $, respectively.
Furthermore, we introduce the bivector representation that consists in defining the curvature components $R_{abcd}$ as the components 
of a $6\times 6$ matrix ${\bf R}_{AB}$ according to the convention proposed in \cite{misner2017gravitation} 
(Chapter 14, Section 14.1, pp. 333-334),  which establishes the following correspondence between tetrad $ab$ and bivector indices $A$:
\begin{align}
01\rightarrow 1\ ,\quad 02\rightarrow 2\ ,\quad 03\rightarrow 3\ ,\quad 23\rightarrow 4\ ,\quad 31\rightarrow 5\ ,\quad 12\rightarrow 6\ .
\end{align}
Hence, 
by using  the symmetries $R_{abcd} = -R_{abdc} =-R_{bacd}$,  the Riemann curvature tensor $R_{abcd}$ can be explicitly expressed as the $6\times 6$ matrix 
\begin{align}
{\bf R}_{AB}=\left(
\begin{array}{cccccc}
 R_{0101} &  R_{0102}  &  R_{0103} &  R_{0123}  & R_{0131} &  R_{0112}   \\
 R_{0201} &  R_{0202}  &  R_{0203} &  R_{0223}  & R_{0231} &  R_{0212}   \\
 R_{0301} &  R_{0302}  &  R_{0303} &  R_{0323}  & R_{0331} &  R_{0312}   \\
 R_{2301} &  R_{2302}  &  R_{2303} &  R_{2323}  & R_{2331} &  R_{2312}   \\
 R_{3101} &  R_{3102}  &  R_{3103} &  R_{3123}  & R_{3131} &  R_{3112}   \\
 R_{1201} &  R_{1202}  &  R_{1203} &  R_{1223}  & R_{1231} &  R_{1212}   
\end{array}
\right),
\label{matexplicit}
\end{align}
Furthermore, due to the symmetry $R_{abcd} = R_{cdab}$, the matrix ${\bf R}_{AB}$ is symmetric with 21 independent components. This number is reduced to 20 by using the algebraic Bianchi identity $R_{a[bcd]}=0$, which in bivector representation reads 
\begin{align}
{\bf R}_{14}+{\bf R}_{25}+{\bf R}_{36}=0\ .
\end{align}
For any given solution of Einstein's equations, the eigenvalues of the Riemann curvature tensor can be calculated in a straightforward way by calculating the eigenvalues of the matrix ${\bf R}_{AB}$ given above.

In general, the matrix $R_{AB}$ can be rewritten in such a way that it contains all the information about the Einstein equations, 
\begin{align}
R_{ab} - \frac{1}{2} R \eta_{ab}  =  k \,T_{ab} \ , \ \  k \equiv 8\pi Gc^{-4} \ ,
\end{align}
with $G$ and $c$  being the Newton gravitational constant and the light speed in the vacuum, respectively.
\footnote{We will use the MKS unit system to analyze the behavior of physical parameters related to observed neutron stars. 
In this system, we consider the values of light speed and Newton's gravitational constant as 
 $c= 299792458$ m/s  and
 $G = 6.674 \times 10^{-11} {\text N} {\text m}^2/ \text{kg}^2 $,
respectively. }
To this end, we write Einstein equations explicitly in terms of the curvature components ${\bf R}_{AB}$, obtaining a set of ten algebraic equations that relate 
the components of ${\bf R}_{AB}$ and $T_{ab}$. 
 Consequently, only ten components of the matrix ${\bf R}_{AB}$ are algebraic independent 
and can be arranged in the $6\times 6$ curvature matrix in the following way
\be \label{eq: CurvatureTensor}
{\bf R}_{AB}=\left(
\begin{array}{cc}
	{\bf M}_1 & {\bf L} \\
	{\bf L} & {\bf M}_2 \\
\end{array}
\right),
\ee
where 
$$
{\bf L} =\left(
\begin{array}{ccc}
{\bf R}_{14}  & 	{\bf R}_{15}  & {\bf R}_{16} \\
{\bf R}_{15} - k T_{03} &  {\bf R}_{25} &  {\bf R}_{26}  \\
{\bf R}_{16} + k T_{02}  &   \quad {\bf R}_{26}  - k  T_{01} & \quad - {\bf R}_{14} - {\bf R}_{25}   \\
\end{array}
\right), 
$$ 
and 
${\bf M}_1$ and  ${\bf M}_2$ are $3\times 3$ symmetric matrices
$$
{\bf M}_1=\left(
\begin{array}{ccc}
{\bf R}_{11} &  \quad	{\bf R}_{12} & {\bf R}_{13} \\
{\bf R}_{12} & \quad {\bf R}_{22} &  {\bf R}_{23} \\
{\bf R}_{13} & \quad   {\bf R}_{23} &  \quad - {\bf R}_{11}   -	{\bf R}_{22}   {+} k \left(\frac{T}{2} +T_{00}\right)  \\
\end{array}
\right),
$$

$$
{\bf M}_2 = 
{   \left( \\
	\begin{array}{ccc}
	-{\bf R}_{11} + k \left(\frac{T}{2} +T_{00}-T_{11} \right)   &  {-} { \bf R}_{12} - k T_{12}   & - {\bf R}_{13} - k T_{13} \\
	{-} { \bf R}_{12} - k T_{12}   &   -{\bf R}_{22} + k \left(\frac{T}{2} +T_{00}-T_{22} \right)   &  - {\bf R}_{23} - k T_{23} \\
	- {\bf R}_{13} - k T_{13}    &    - {\bf R}_{23} - k T_{23}   &   {\bf R}_{11} +	{\bf R}_{22}  {-} k T_{33}   \\
	\end{array}
	\right)   },
$$     
with $T=\eta^{ab}T_{ab}$. 
This is the most general form {of} a curvature tensor that satisfies Einstein's equations with an arbitrary energy-momentum tensor.
The eigenvalues $\lambda_n\ (n=1,\cdots,6)$ of the matrix ${\bf R}_{AB}$ are known as the curvature eigenvalues. 
It is convenient to express the eigenvalues $\lambda_n$  in terms of the components of the Riemann tensor $R_{abcd}$. To do this, 
we consider the simplest case in which the curvature matrix ${\bf R}_{AB}$  is diagonal. Then, from the explicit form of the curvature, 
 matrix (\ref{matexplicit}), it follows that 
\begin{align}
\lambda_1 = R_{0101}\ , \ \lambda_2 = R_{0202}\ , \ etc.
\end{align}
i.e., the eigenvalues coincide with the diagonal components of ${\bf R}_{AB}$. This shows that the eigenvalues are just the non-zero tetrad
 components of the curvature tensor. In general, the eigenvalues depend only on the tetrad components and can be expressed as rational 
 functions in which the order of the polynomials depends on the number of non-zero tetrad components. 

An important property of the eigenvalues is that they characterize uniquely a given spacetime. Indeed, given the metric ${\cal G}$, 
the calculation of the eigenvalues $\lambda_n$ does not allow any arbitrariness. As the curvature is a measure of the gravitational interaction, 
we conclude that the eigenvalues $\lambda_n$ should contain all the information about the behavior of the gravitational interaction of a given spacetime metric.

Consider now two spacetimes $({\cal M^+, G^+})$ and $({\cal M^-, G^-})$ that are separated by a hypersurface $\Sigma$. Consequently, each spacetime is characterized by a unique set of curvature eigenvalues, say $\lambda_n^+$ and $\lambda_n^-$, respectively. 
The matching problem consists in ``gluing" these two spacetimes along the hypersurface $\Sigma$ in such a way that the resulting differential manifold is correctly defined and describes a gravitational field. The solution of this matching problem offered by the $C^3$ approach consists in demanding 
that the eigenvalues coincide across the matching hypersurface, i.e., 
\be
\lambda_n ^+|_\Sigma = \lambda_n ^-|_\Sigma\ , \quad \forall n\ .
\ee
This simple condition essentially means that the curvature is continuous along the matching surface. 

In the case of astrophysical compact objects, which we will now consider, we identify $({\cal M^+, G^+})$ and 
$({\cal M^-, G^-})$  as representing the exterior and interior gravitational fields of the object, respectively. 
Consequently, the hypersurface $\Sigma$ can be identified with the surface of the object.
The main advantage of the $C^3$ approach is that the curvature is represented in an invariant way through the eigenvalues, implying that the results are coordinate independent.  
A second advantage of the $C^3$ matching approach is that it is necessary to specify matching surface $\Sigma$ {\it a priori}; instead, it is determined by the matching 
radius, $r_{match} $, defined as 
\be 
r_{match} \in [r_{rep}, \infty) \ ,  \quad r_{rep} ={\rm max}\{r_l\} \  ,  
\label{rmatch}                 
\ee
where $r_l$  $(l = 1,2,... )$, with $0 < r_l < \infty$,    represents the set of solutions of the equation
\be
\frac{\partial\lambda_n^+}{\partial r}\Big|_{r=r_l} = 0 \ .
\ee
In this work, we assume  
that the manifold $({\cal M}^+, {\mathbf{\cal G}}^+)$ is  asymptotically 
flat, i.e., there exists a spatial coordinate $r$ such that
 \be
 \lim_{r\to\infty} {\mathbf{\cal G}}^+ = {\bf \eta}
 \ee
 where ${\bf \eta}$ represents the Minkowski metric.

 The condition (\ref{rmatch}) is defined in terms of the repulsion radius $r_{rep}$, which is defined as the location where the first local extremum is found in an eigenvalue as approaching the object from infinity, i.e., the location starting from which repulsive gravity could be detected.  Then, from a physical point of view condition (\ref{rmatch}) 
 means that  the matching surface is placed so that no repulsive gravity is present.


\section{$C^3$ matching for anisotropic fluids}
\label{sec:sph}

In this section, we will apply the $C^3$ matching approach to match a spherically symmetric spacetime describing an anisotropic fluid 
 $({\cal M}^-, {\mathbf{\cal G}}^-)$ to  an asymptotically flat spacetime $({\cal M}^+, {\mathbf{\cal G}}^+)$, which satisfy Einstein's equations. 
In the interior and exterior regions, we choose spherical coordinates and metrics of the form 
      \begin{align}\label{eq:Line_Element}
{\mathbf{\cal G }} =   
- e^{\nu}c^2    
\operatorname{d}t \otimes    
\operatorname{d}t    
+ e^{\phi}   \operatorname{d}r \otimes    
\operatorname{d}r  
+  r^2 \operatorname{d}\Omega \otimes    \operatorname{d} \Omega \, 
\end{align}
where 
       $ \operatorname{d}\Omega \otimes    \operatorname{d} \Omega \equiv   \operatorname{d}\theta \otimes   \operatorname{d}\theta   
          + \sin^2\theta     \operatorname{d}\varphi \otimes    \operatorname{d}\varphi $
and the functions  $\nu$ and $\phi$  depend  on $r$ only.  Similarly, we suppose that the conventional matter governing the internal spacetime 
dynamics is a fluid determined by the energy-momentum tensor  
     \begin{align}\label{eq:AnisotropicFluidT1}
       T^{\alpha \beta} = (\mu c^2 + p_1)  \vartheta_t^{ \ \alpha}   \vartheta_t^{\ \beta}  + p_1 {\mathbf {\cal G}}^{\alpha \beta}
       + (p_2  -  p_1)  \vartheta_{\theta}^{ \ \alpha}   \vartheta_{\theta}^{\ \beta}
       + (p_3  -   p_1)  \vartheta_{\varphi}^{ \ \alpha}   \vartheta_{\varphi}^{\ \beta} \ ,
        \end{align}
here   $\vartheta_t^{ \ \alpha} $    is the four-velocity of the fluid,  $\mu$ is the volumetric mass density, $(p_1, p_2, p_3)$ are the components of the 
pressure and the basis  $(\vartheta_t, \vartheta_r, \vartheta_{\theta}, \vartheta_{\varphi})$  is dual to the one-forms 
               \begin{align}\label{eq:one-forms}
                \vartheta^t             =  e^{\nu/2}c \operatorname{d}t,  \quad
                \vartheta^r             =  e^{\phi/2}   \operatorname{d}r \ , \quad
                \vartheta^{\theta}   =  r \operatorname{d} \theta \ , \quad
                \vartheta^{\varphi}      =  r \sin{\theta}\operatorname{d}\varphi \ .
                  \end{align}    
 Using the expression for the energy-momentum tensor corresponding to an anisotropic fluid (\ref{eq:AnisotropicFluidT1}), a direct 
 computation  shows that  for the interior  spacetime  $({\cal M}^-, {\mathbf{\cal G}}^-)$, the curvature matrix  ${\bf R}_{AB}$ is diagonal 
 and, according to Eq.(\ref{eq: CurvatureTensor}), the eigenvalues are
     \begin{align}
      \label{eq: AnisFluid_Eigenvalues_1}
      \lambda^-_1 & = R_{0101} = \frac{e^{-\phi}}{4}(2\nu_{,rr}+\nu_{,r}^2 -\nu_{,r}\phi_{,r})  \ ,\\
      \lambda^-_2 & = R_{0202}  = \frac{e^{-\phi} \nu_{,r}}{2 r} \ , \\
      \lambda^-_3 & =   - \lambda^-_1 - \lambda^-_2   + \frac{4\pi G}{c^4} (c^2\mu + p_1 + p_2 + p_3)   \ ,\\
      \lambda^-_4 & =   - \lambda^-_1   + \frac{4\pi G}{c^4} (c^2\mu -  p_1 + p_2 + p_3)  \ , \\
      \lambda^-_5 & =   - \lambda^-_2   + \frac{4\pi G}{c^4} (c^2\mu +  p_1)  \ , \\
      \lambda^-_6 & = \lambda^-_1 + \lambda^-_2 -  \frac{8\pi G}{c^4}  p_2  \ .
      \label{eq: AnisFluid_Eigenvalues_6}
         \end{align}    
On the other hand, according to Birkhoff's theorem, the exterior spacetime $({\cal M}^+, {\mathbf{\cal G}}^+)$ must be described 
by the Schwarzschild metric
      \begin{align}\label{eq: Schwarzschild_Solution}
        {\mathbf{\cal G}}^+ =   - \left(1 -  \frac{2 M G}{c^2r} \right) c^2   \operatorname{d}t \otimes  \operatorname{d}t    
                                        + \left(1 -  \frac{2 M G}{c^2 r} \right) ^{-1}   \operatorname{d}r \otimes \operatorname{d}r    
                                        + r^2 \operatorname{d}\Omega \otimes    \operatorname{d} \Omega .
          \end{align}  
A straightforward computation shows that  for the exterior spacetime, the curvature matrix  ${\bf R}_{AB}$ is diagonal 
and the eigenvalues are
    \begin{align}
     \lambda^+_2  =   \lambda^+_3 =   - \lambda^+_5=  - \lambda^+_6 =  \frac{G M}{c^2r^3}\ , \qquad
     \lambda^+_1  =  - \lambda^+_4 =  - \frac{ 2 GM}{c^2r^3} \ . 
     \label{eq: Schwarzschild_Eigenvalues_6}
     \end{align}     
The spacetimes  $({\cal M}^-, {\mathbf{\cal G}}^-)$  and  $({\cal M}^+, {\mathbf{\cal G}}^+)$ can be matched 
at the surface $\Sigma$, determined by the matching radius $r_{match}$  as defined in Eq.(\ref{rmatch}),
 if the  necessary and sufficient condition
 $\lambda_n^-|_\Sigma - \lambda_n ^+|_\Sigma = 0 ,\quad n=1,\cdots, 6
$
is satisfied. Using the above expressions for the eigenvalues, we obtain that the following system of algebraic equations must be satisfied at the matching surface   $\Sigma$,
        \begin{align}
         c^2 \mu + p_1 + p_2 + p_3 =0,\\
         c^2 \mu - p_1 + p_2 + p_3 =0,\\
         c^2 \mu + p_1 =0,  \\
         p_2 = 0, \\
         p_2 = p_3,
         \end{align}
whose only solution is  $ c^2 \mu = p_1 = p_2  = p_3 =0$. This result implies that the density and pressures of the compact object should vanish at the surface in order for the matching conditions to be satisfied. From a physical point of view, this is an expected result since the interior anisotropic fluid cannot be part of the exterior vacuum spacetime. 


\section{Discontinuous matching}
\label{sec:disc}

In this section, we formulate a generalization of the procedure presented in the previous section, which  allows us to consider the case of solutions having 
 non-zero density and anisotropic pressures on the matching surface.
 This means that the eigenvalues could be discontinuous on the matching surface, i.e.,  $\lambda_n^+\neq \lambda_n ^-$ on $\Sigma$ for at least one value of $n$. 

To formulate the $C^3$ matching conditions in the case of an anisotropic fluid with discontinuities, we will follow Israel's thin-shell approach \cite{israel1966singular} and the $C^3$ discontinuous matching for perfect fluids \cite{gutierrez2022darmois}. 
 To this end, let us consider the jump of the eigenvalues 
across $\Sigma$ as 
\be
[\lambda_n] = \lambda_n ^- - \lambda_n ^+ \ ,
\ee
and the 
jump of the Einstein tensor and the energy-momentum tensor along  $\Sigma$, i.e., 
\be
[ G_{ij}] =  G_{ij}^- - G_{ij}^+\ , \quad
[ T_{ij}] =  T_{ij}^- - T_{ij}^+\ , 
\ee
with 
\be
 G_{ij} ^\pm = \frac{\partial x^{\alpha}_\pm}{\partial \xi^i}
\frac{\partial x^{\beta}_\pm}{\partial \xi^j}  G_{\alpha\beta}^\pm\ , \quad
 T_{ij} ^\pm = \frac{\partial x^{\alpha}_\pm}{\partial \xi^i}
\frac{\partial x^{\beta}_\pm}{\partial \xi^j} T_{\alpha\beta}^\pm\ ,
\ee
where $\xi^i$ are the coordinates of the surface $\Sigma$ and $x^\mu_\pm$ are the coordinates of the interior and exterior spacetimes,  respectively.
Furthermore, the jump of the Einstein tensor is used to define the 
 energy-momentum tensor of the shell ${ S}_{ij}$ as
 \be 
[G_{ij}] = \frac{8 \pi G}{c^4}  S_{ij}\ .
\label{ie1}
\ee
To guarantee that ${S_{ij}}$ describes the energy-momentum tensor of a realistic thin shell, we demand that the components of  $ S_{ij}$ be induced 
by the energy-momentum tensors of the interior and exterior spacetimes as follows  
\begin{align}\label{eq:AnisotropicFluidT2}
 S^{i j} = [T^{ij}] =  (\sigma c^2  + P_1)  \vartheta_0^{ \ i}   \vartheta_0^{\ j}  
+ P_1 {\mathbf {\cal \gamma}}^{i j}
+ (P_2  -  P_1)  \vartheta_{2 }^{ \ j}   
\vartheta_{2}^{\ j}
+ (P_3  -   P_1)  \vartheta_{3}^{ \ \alpha}   \vartheta_{3}^{\ j} \ ,
\end{align}
where $\sigma$,  $P_1$,  $P_2$, and $P_3$  are the energy density and the anisotropic pressures of the fluid evaluated at the matching surfaces, i.e., 
\be
\sigma = \mu|_\Sigma\ , \qquad P_1= p_1 |_\Sigma\  , \qquad P_2 = p_2 |_\Sigma\  ,  \qquad P_3= p_3 |_\Sigma\  .
\label{eq:bound}
\ee

In summary, in  case of discontinuities, we will say that an interior spacetime can be matched with an exterior one along a boundary shell 
located on $\Sigma$,  if there exists a density $\sigma$ and pressures $P_1$, $P_2$, $P_3$ satisfying the induced Einstein equations 
(\ref{ie1}) and (\ref{eq:AnisotropicFluidT2}).

Furthermore, the jumps of the eigenvalues will depend on the explicit form of the solutions. For a general interior solution, using the explicit expressions for the eigenvalues presented in Eqs.(\ref{eq: AnisFluid_Eigenvalues_1})--(\ref{eq: AnisFluid_Eigenvalues_6}) and (\ref{eq: Schwarzschild_Eigenvalues_6}), and defining the matching surface as a sphere of radius $r_{match} = R$, we obtain
\begin{align}
[\lambda_1]  = &   
\lambda_1^-|_\Sigma + \frac{2GM}{c^2 R^3} ,  \\ 
[\lambda_2]  = &   
\lambda_2^-|_\Sigma - \frac{GM}{c^2 R^3} ,\\
[\lambda_3]  = & 
-\lambda_1^-|_\Sigma - \lambda_2^-|_\Sigma 
+ \frac{4\pi G}{c^4}(c^2 \sigma + P_1+P_2+P_3) -\frac{GM}{c^2 R^3} 
\ , \\
[\lambda_4]  = &
- \lambda_1^-|_\Sigma 
+ \frac{4\pi G}{c^4}(c^2 \sigma - P_1+P_2+P_3) -\frac{2GM}{c^2 R^3} 
\ , \\
[\lambda_5]  = & 
- \lambda_s^-|_\Sigma 
+ \frac{4\pi G}{c^4}(c^2 \sigma + P_1) +\frac{GM}{c^2 R^3} 
\ , \\
[\lambda_6]  = &
\lambda_1^-|_\Sigma + \lambda_2^-|_\Sigma 
- \frac{8\pi G}{c^4} P_2 +\frac{GM}{c^2 R^3}
\ ,
\label{eq:EigenvaluesJump1}
\end{align}
where
\be
\lambda_1^-|_\Sigma = 
\frac{e^{-\phi}}{4}(2\nu_{,rr}+\nu_{,r}^2 -\nu_{,r}\phi_{,r}) \Big |_{r=R}, \quad
\lambda_2^-|_\Sigma =
\frac{e^{-\phi} \nu_{,r}}{2 r}\Big|_{r=R} \ .
\ee

In the following section, we will apply the matching procedure described above  to a particular interior solution. 

\section{$C^3$ matching conditions for neutron stars}
\label{sec:C3NS}

 In this section, we will examine a particular class of static spherically symmetric solutions of Einstein equations 
solutions, which can be used to describe the interior gravitational field of neutron stars.   Consider the solutions recently presented 
in  \cite{solanki2021new}  by Solanki and Takore (ST), which describe the spacetime interior region $({\cal M}^-, {\mathbf{\cal G}}^-)$ with the line element 	
\begin{align}\label{eq:Int_Line_Element}
{\mathbf{\cal G }} =   -\frac{ \left( 1 + \frac{r^2}{\beta^2}\right)^{\alpha}}
{\left ( 1 - \frac{\kappa r^2}{\beta^2} \right)^{\frac{1+\kappa}{2\kappa}}e^{\frac{\alpha r^2}{2 \beta^2}}  }
c^2 \operatorname{d}t \otimes    \operatorname{d}t     + \frac{\left( 1 + \frac{r^2}{\beta^2}\right)}{  1 - \frac{\kappa r^2}{\beta^2}}  \operatorname{d}r \otimes \operatorname{d}r  
   +  r^2 \operatorname{d}\Omega \otimes    \operatorname{d} \Omega \, 
\end{align}
where   $\alpha$, $\beta$ and $\kappa$ are arbitrary parameters. 
The nature of the spacetime is determined by the energy-momentum tensor (\ref{eq:AnisotropicFluidT1}), 
where the  mass density reads
	 \begin{align}\label{eq:EnergyDensity_ST}
	 \mu = \frac{c^2}{8 \pi G}  \frac{\left(\frac{1+ \kappa}{\beta^2}\right)(3 + \frac{r^2}{\beta^2})}{ \left( 1 +   \frac{r^2}{\beta^2}  \right)^2 } \ ,
	 \end{align}	
and the components of the pressure are
	 \begin{align}\label{eq:P1_ST}
	 p_1 = \frac{c^4}{8 \pi G}  \frac{   {\alpha} \left( 1 - \frac{\kappa r^2}{\beta^2}  \right)  
	              \left(   1 - \frac{ r^2}{\beta^2}  \right) }{ {\beta^2} \left( 1 +   \frac{r^2}{\beta^2}  \right)^2 } \ ,
	 	 \end{align}
and 
   \begin{align}\label{eq:P2_ST}
	 p_2 = p_3 & =  \frac{c^4 H}{8 \pi G \beta^2   (  1    +  \frac{r^2}{\beta^2})^3 }  \ , \\
	             H & =   \frac{1}{ 4 (1  - \frac{\kappa r^2}{ \beta^2}   )} 
	                   \Big\{ 
	                     4\alpha   +  \left[  \alpha^2 - 8 (\kappa + 1) \alpha + 3(\kappa + 1)^2  \right] \frac{r^2}{\beta^2} \nonumber \\
	               & - 2 \left[  (\kappa + 1)\alpha^2 - ( 2 \kappa^2 + 9 \kappa - 1)\alpha  - 2 (\kappa + 1)^2  \right] \frac{r^4}{\beta^4} \nonumber\\
                       &  +  \left[   ( \kappa^2 + 4\kappa + 1)\alpha^2 - 2( 5 \kappa^2 - 2\kappa + 1) \alpha  + (1 + \kappa)^2 \right] \frac{r^6}{\beta^6}\nonumber\\
                       & - 2 \alpha \kappa \left[   (1+ \kappa) \alpha + \kappa - 1 \right] \frac{r^8}{\beta^8}  + \alpha^2 \kappa^2 \frac{r^{10}}{\beta^{10}}
	                      \Big\} \ . 
	                     \nonumber
	 	 \end{align}
We will match this interior solution with the exterior spacetime $({\cal M}^+, {\mathbf{\cal G}}^+)$ described 
by the Schwarzschild metric  (\ref{eq: Schwarzschild_Solution}). We will see  how the matching conditions  determine the values of the free parameters.

A computation reveals that the curvature matrix ${\bf R}_{AB}$ is diagonal for the interior spacetime, with eigenvalues:
           \begin{align}
            \lambda^-_{1} &= \frac{L}{4\beta^4(\beta^2 + r^2)^3(\kappa r^2 - \beta^2)} \ , \\
                          L  & \equiv  - 2 (\alpha + \kappa + 1)\beta^{10} - [ \kappa^2  - 4(\alpha - 1) \kappa + \alpha^2 - 8\alpha + 3] \beta^8 r^2  \nonumber\\  
                                  & + 2 [  - (\alpha + 2)\kappa^2  + (\alpha^2 - 9 \alpha - 3)\kappa + \alpha^2 - 1 ]\beta^6 r^4\nonumber\\  
                                  &  - [ ( \alpha^2  - 10 \alpha + 3) \kappa^2 + 4 (\alpha^2 + 1)\kappa + (\alpha - 1)^2] \beta^4 r^6  \nonumber\\  
                                  &  -   2 \alpha \kappa (\alpha \kappa + \alpha - 1)\beta^2 r^8 - \alpha^2 \kappa^2 r^{10} \ , \nonumber                    
              \end{align}    
          \begin{align}
            \lambda^-_{2}  & =  \lambda^-_{3} =   \frac{(\alpha + \kappa + 1)\beta^4 - (1 + \kappa)(\alpha - 1)\beta^2 r^2 
            + \alpha\kappa r^4}{2\beta^2(\beta^2 + r^2)^2}  \ , \nonumber
              \end{align}	
          \begin{align}
            \lambda^-_{4} = \frac{1 + \kappa}{ \beta^2 + r^2} \ ,      \quad
            \lambda^-_{5} =    \lambda^-_{6} =  \frac{(1 + \kappa)\beta^2}{( \beta^2 + r^2)^2} \ .       \nonumber
              \end{align}  	
In order to satisfy the matching conditions (\ref{ie1}) and (\ref{eq:AnisotropicFluidT2}), 
it is convenient to fix the matching radius in terms of the parameters entering the metric as 
 \begin{align}
  r_{match} = \beta = \frac{4GM}{(1 + \kappa)\ c^2} \ ,  \qquad \alpha = \frac{(\kappa + 1)^2}{(\kappa - 1)^2}.
  \label{rmatch}
   \end{align}
Then,  the jump in the eigenvalues reads
 \begin{align}
 [\lambda_1] & = [\lambda_5] =[\lambda_6] = \frac{(1+\kappa)^3 c^4}{32G^2M^2}  = \frac{4 \pi G}{c^2}\sigma \ , \\ 
[\lambda_2 ] & = [\lambda_3] = [\lambda_4] =0 . \nonumber
   \end{align}
Furthermore,  the jump of the induced Einstein tensor reads
 \begin{align}
 [ G_{tt}]  = \frac{(1+\kappa)^3 c^4}{16G^2M^2} , \qquad    [G_{\theta \theta}] =  [G_{\varphi \varphi}] = 0 \ .
   \end{align}    
Therefore,  the induced Einstein equations 
   \begin{align}
[ G_{ij}] = \frac{8 \pi G}{c^4}   S_{ij}
 \end{align}
are satisfied for 
\begin{align}
 S_{ij} =c^2 \sigma  U_iU_j \ ,  \qquad
  \sigma= \frac{(1+\kappa)^3 c^6}{128 \pi G^3M^2}  \ , \qquad U_{i} = (-1,  0,  0) \ .
   \end{align}
 We conclude that the discontinuous matching conditions (\ref{ie1}) are satisfied, and the matching surface $\Sigma$ corresponds to a dust thin shell of radius $r_{match}$ and mass density $\sigma$. The matching radius and the surface density of the thin shell are entirely given in terms of the parameter $\kappa$, which characterizes the interior spacetime, and the mass parameter $M$ of the exterior Schwarzschild spacetime.
  
\begin{figure}[h!]
	\centering
	 { \includegraphics[width = 0.48\textwidth]{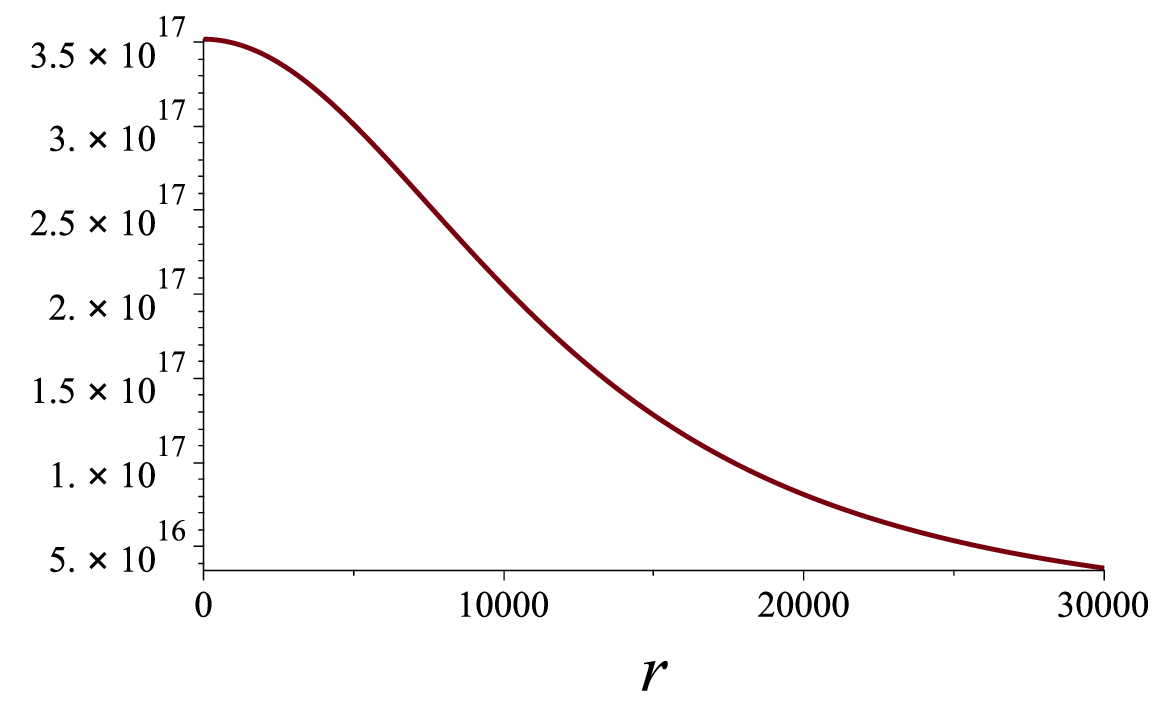}  }
	 { \includegraphics[width = 0.48\textwidth]{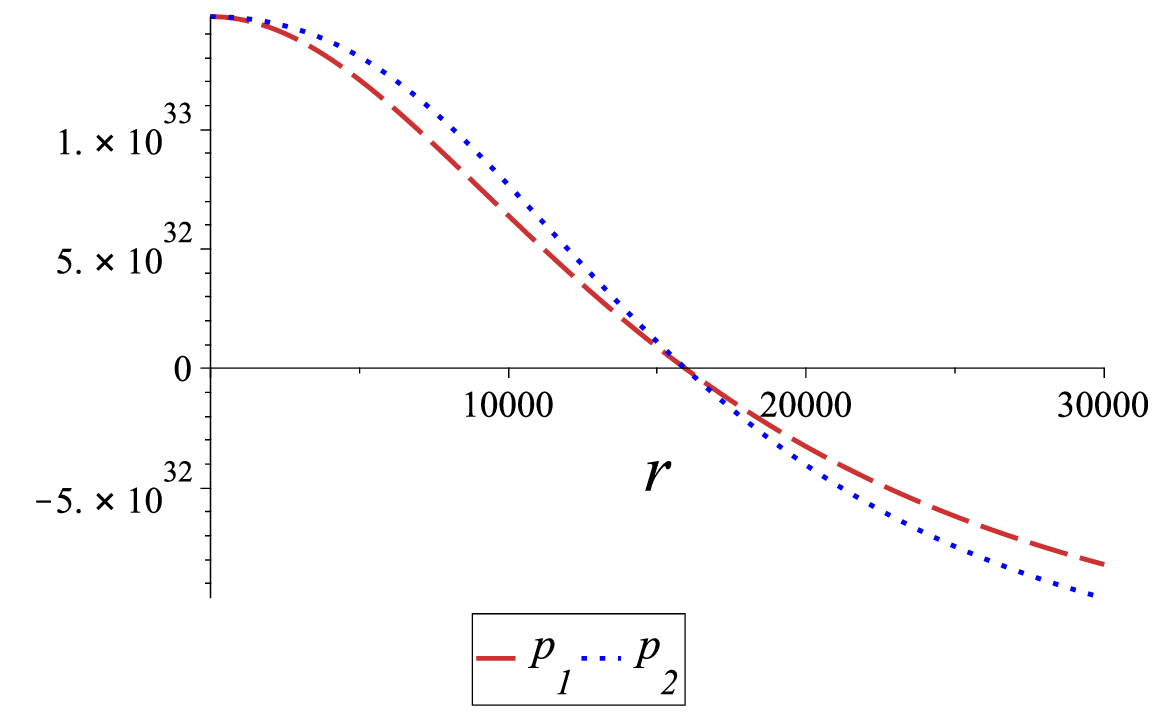} } 
	 \caption{ The mass density $\mu$ (in kg/m$^3$) and pressures $p_1$ and $p_2$ (in Pascals) for the interior ST metric 
for $\kappa = -0.9944404128$ and $M= 1.5 {\text M}_{\odot} $. The radius of the star is 15940m.   }
\label{fig:MassPress}
\end{figure}

\begin{figure}[h!]
	\centering
	\subfloat[$\lambda_1$]{  \includegraphics[width = 0.5\textwidth]{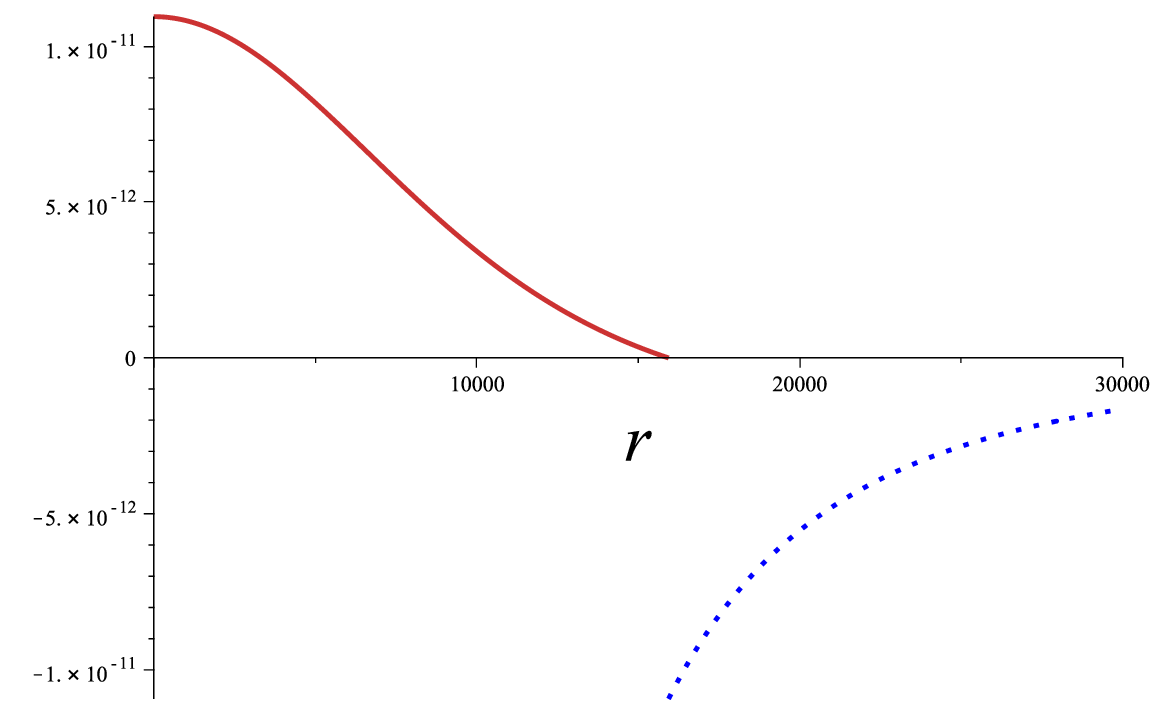} }
	\subfloat[$\lambda_2$]{  \includegraphics[width = 0.5\textwidth]{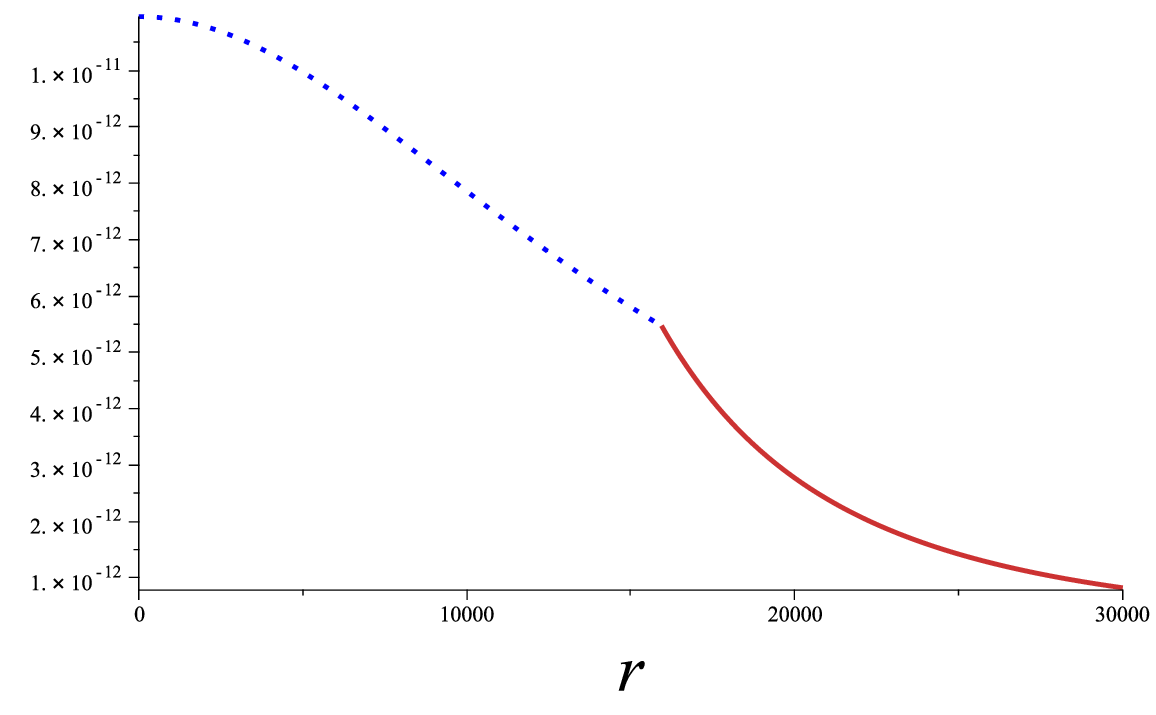} } \\
\subfloat[$\lambda_3$]{  \includegraphics[width = 0.5\textwidth]{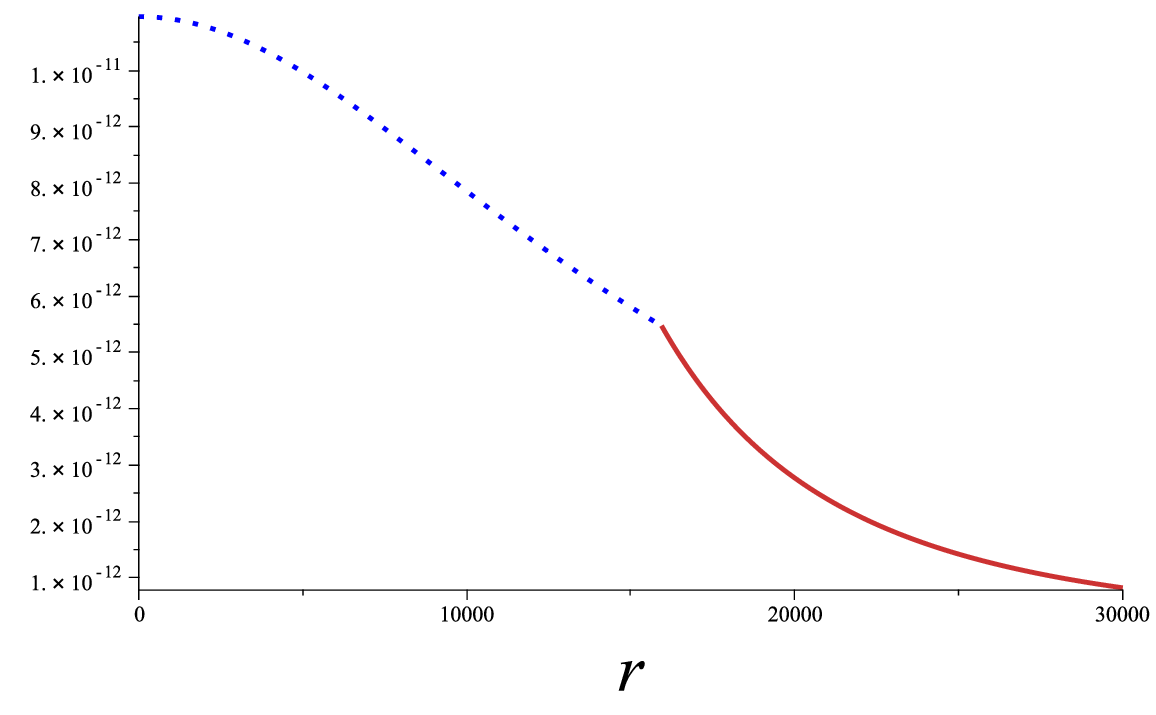} }
\subfloat[$\lambda_4$]{  \includegraphics[width = 0.5\textwidth]{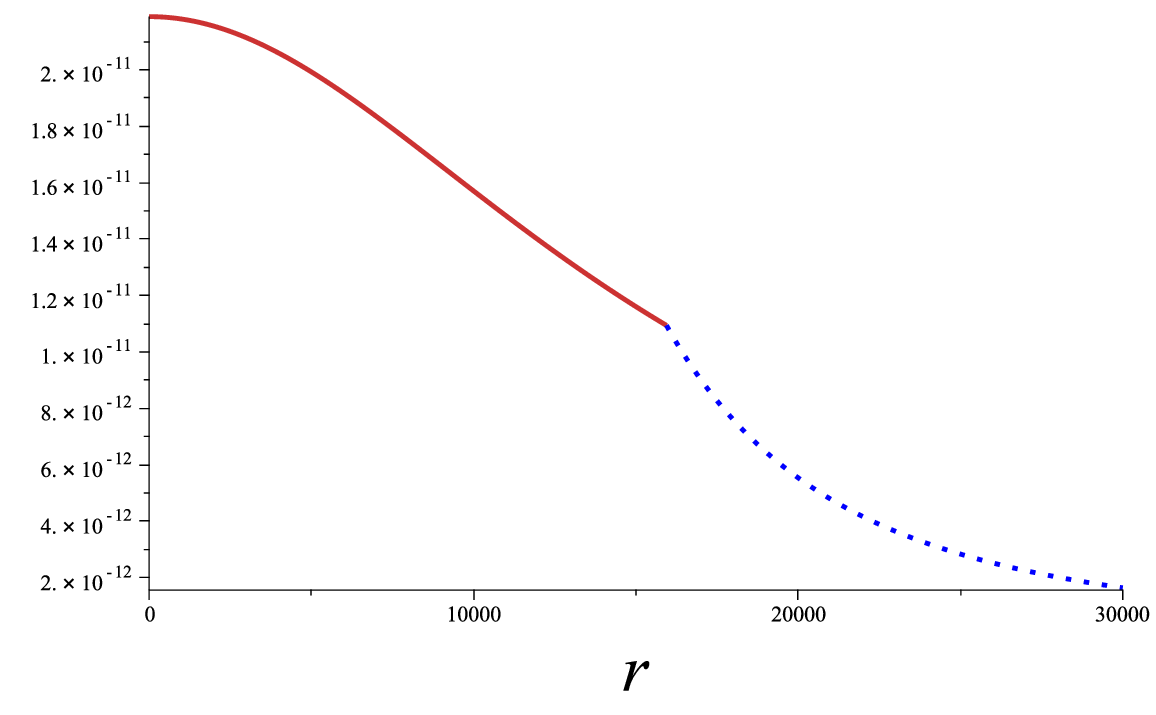} } \\
\subfloat[$\lambda_5$]{  \includegraphics[width = 0.5\textwidth]{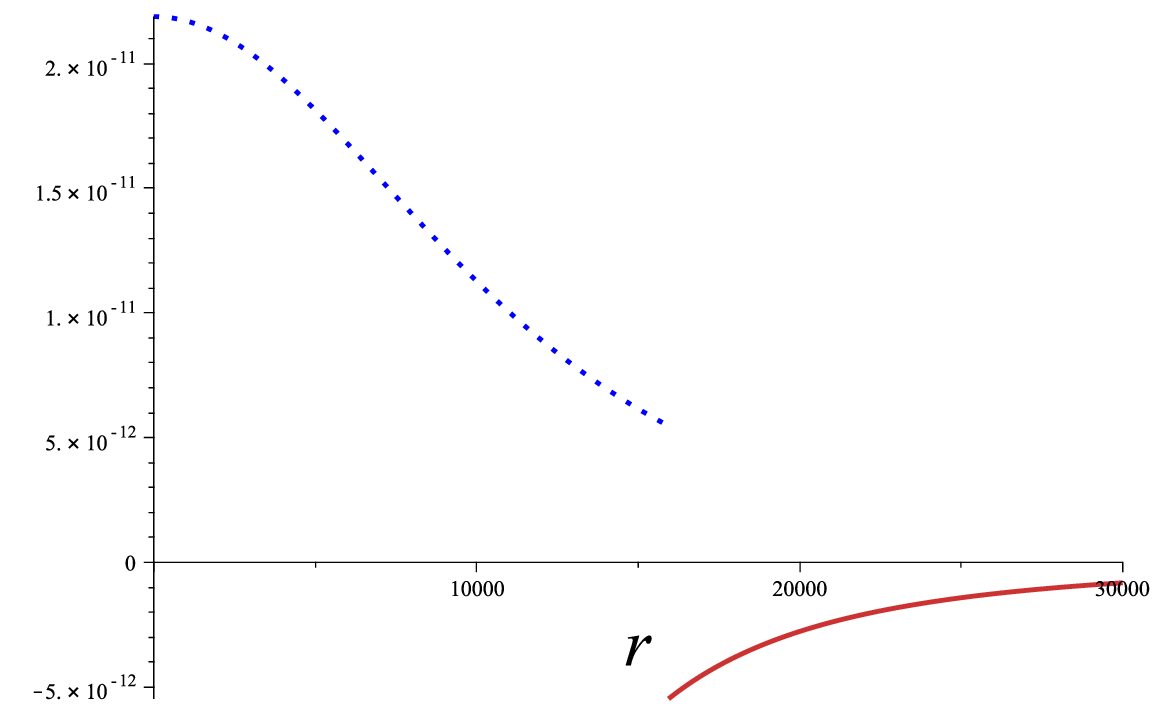} }
\subfloat[$\lambda_6$]{  \includegraphics[width = 0.5\textwidth]{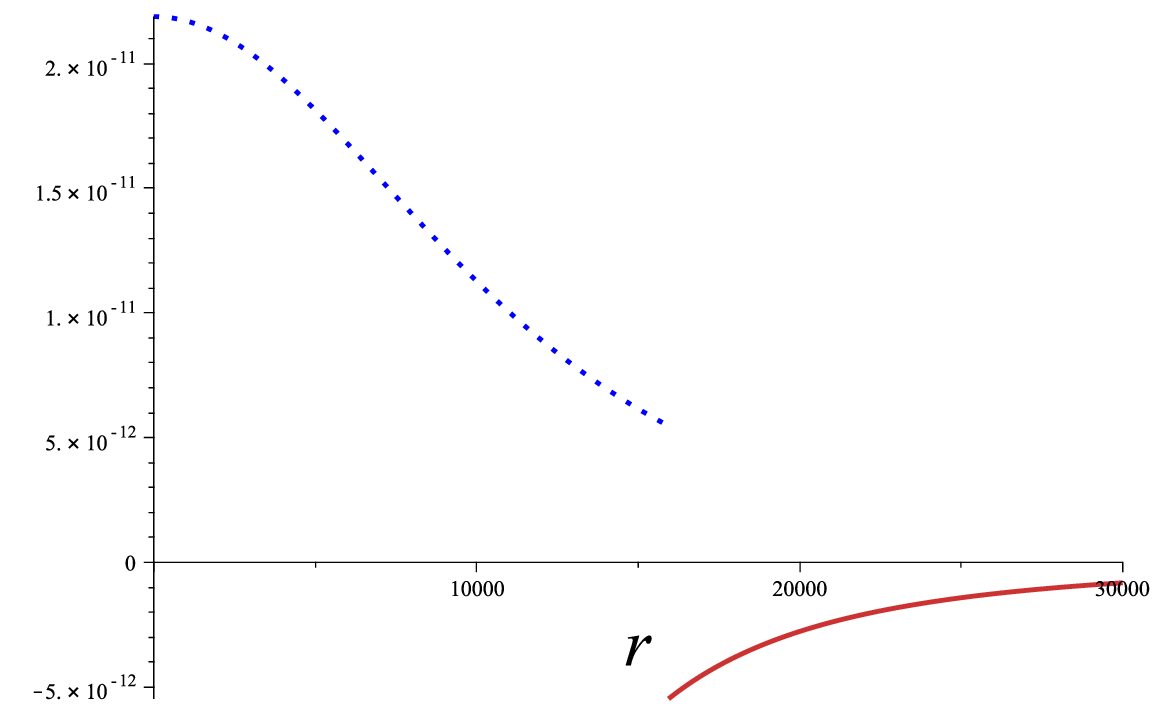} } 	
\caption{ The curvature eigenvalues (in $\text m^{-2}$)for the interior ST metric ($r \leq  15940$m ) 
and the exterior Schwarzschild metric ($r\geq 15940$m ) for $\kappa = -0.9944404128$ and $M= 1.5 {\text M}_{\odot} $.}
\label{fig:Eigenvalues}
\end{figure}

To demonstrate the practical application of the $C^3$ matching formalism, let us consider the case of a neutron star of mass  
1.5 solar masses (${\text M}_{\odot}  = 1.989 \times 10^{30} {\text kg} $) and a radius of 15940m, which we identify with the matching radius (\ref{rmatch}). Then, the value of the internal parameter is  $\kappa =-0.9944404128$. 
Thus, we see that given a mass and a radius for the neutron star, we can find the compatible value for the free parameter $\kappa$. 
 We have tested  various  $(\kappa, {\text M})$ values  for compatible $r_{match}$ values. Our results are in agreement with the values presented 
 in \cite{solanki2021new}, which have been derived  for realistic neutron star configurations. 
  In Fig. \ref{fig:MassPress}, we illustrate the behavior of the mass density and the anisotropic pressures for the above example of a neutron star. The pressures vanish at the radius $r_{match}=15940$m, but the mass density is different from zero, indicating the presence of a discontinuity.
  In Fig.  \ref{fig:Eigenvalues}, we show the behavior of the corresponding curvature eigenvalues inside and outside the star, indicating in each case the discontinuities located at the surface of the star. 

\section{Conclusions}
\label{sec:CR}

The $C^3$ procedure has been proposed recently as an invariant approach to matching spacetimes along a specific matching surface. The advantage of the $C^3$ approach is that it is based upon the use of scalar quantities represented by eigenvalues of the Riemann curvature tensor. Moreover, it allows us to determine the position of the matching surface by using the behavior of the curvature eigenvalues as the source of gravity is approached from infinity.

In this work, we applied the $C^3$ approach to match an interior solution of Einstein's equations, with an energy-momentum tensor representing an anisotropic fluid, with the exterior Schwarzschild spacetime. The main result in this case is that the matching conditions are satisfied only if the mass density and pressures of the anisotropic fluids vanish on the matching surface. This result agrees with our physical expectations since for a smooth transition from the interior spacetime to the exterior vacuum spacetime, the fluid should vanish at the matching surface.

We generalized the $C^3$ approach to include the case in which the mass density and the pressures of the fluid do not vanish on the matching surface. This implies that discontinuities can appear in the physical parameters of the fluid. We use the thin-shell method, which consists in interpreting the discontinuities as due to the presence of an additional fluid that covers the matching surface. To this end, we essentially demand that the parameters of the additional fluid be determined by the discontinuities of the energy-momentum tensor of the interior anisotropic fluid. As a result, we obtain matching conditions, which  guarantee that the additional fluid is physically meaningful.

We tested the $C^3$ discontinuous matching approach in the case of a particular exact interior solution, imposing  values for the total mass and radius of a compact object  that correspond to realistic neutron stars. As a general result, we obtained that the $C^3$ matching conditions can be used to determine the properties of spacetimes, which describe the interior as well as the exterior gravitational field of neutron stars. 

An important assumption of the analysis described in this work is the spherical symmetry, which implies that the gravitational source is static. A more realistic analysis should take into  account the rotation of the source. We expect to investigate this case in future works.  


\end{document}